# Revolutionizing Mental Health Care through LangChain: A Journey with a Large Language Model

Aditi Singh
Computer Science
Cleveland State University
a.singh22@csuohio.edu

Abul Ehtesham
The Davey Tree Expert Company
abul.ehtesham@davey.com

Saifuddin Mahmud
Computer Science & Information Systems
Bradley University
smahmud@bradley.edu

Jong-Hoon Kim*
Computer Science,
Kent State University,
jkim72@kent.edu

*Abstract*— Mental health challenges are on the rise in our modern society, and the imperative to address mental disorders, especially regarding anxiety, depression, and suicidal thoughts, underscores the need for effective interventions. This paper delves into the application of recent advancements in pretrained contextualized language models to introduce MindGuide, an innovative chatbot serving as a mental health assistant for individuals seeking guidance and support in these critical areas. MindGuide leverages the capabilities of LangChain and its ChatModels, specifically ChatOpenAI, as the bedrock of its reasoning engine. The system incorporates key features such as LangChain's ChatPrompt Template, HumanMessage Prompt Template, ConversationBufferMemory, and LLMChain, creating an advanced solution for early detection and comprehensive support within the field of mental health. Additionally, the paper discusses the implementation of Streamlit to enhance the user experience and interaction with the chatbot. This novel approach holds great promise for proactive mental health intervention and assistance.

## I. INTRODUCTION

The issue of mental health is an international situation, affecting people in each particularly developed nations and emerging markets. According to the World Health Organization's Mental Health Action Plan (2013-2020), it's far estimated that around one in four humans international face numerous kinds of mental disorders. This statistic underscores the vast nature of mental health demanding situations throughout extraordinary demographic businesses and areas.

However, what makes this situation even extra complex is the concerning truth that three out of each four people dealing with severe intellectual disorders do no longer have get entry to the necessary remedy they require. This remedy gap intensifies the weight of intellectual health troubles, leaving a sizable part of the populace without the assist and care needed to efficiently address their intellectual health issues.

Furthermore, periods like the recent global pandemic, the effect of mental health issues becomes even more said. The COVID-19 pandemic, in particular, has highlighted how public health crises can extensively have an effect on mental properly-being. During such hard instances, a widespread part of the population faces extended problems in having access to mental fitness professionals. This emphasizes the urgent want for progressed intellectual health offerings and support structures. It underscores the urgency of addressing the mental health disaster and developing complete answers to make certain that people global have the means to successfully deal with their mental fitness challenges..

In studies [1], it's pretty clear that there's a deep connection between mental troubles and the chances of someone taking their own life. And when you look at the big picture, it's quite shocking - nearly a million people across the globe end their lives every year, especially the young ones, making it the second biggest reason for their passing. It's intriguing that when someone attempts suicide, they often grapple with mental challenges. It's like shifting from struggling with difficult thoughts to considering ending everything. This shift is observable in how people express themselves and interact[2].

One practical approach to addressing mental illness and preventing suicidal ideation is early identification. Recent advancements in deep learning have facilitated the development of effective early detection methods [3]. A notable trend in natural language processing (NLP) involves the use of contextualized pretrained language models [4], which have garnered substantial attention for their effectiveness in various text processing tasks.

This paper delves into the application of these recent advancements in pretrained contextualized large language models to introduce MindGuide, an innovative chatbot designed to function as a mental health assistant for individuals in need of guidance and support in these critical areas. MindGuide relies on the capabilities of LangChain and its ChatModels [5], specifically ChatOpenAI [6], as the foundation of its reasoning engine. The system incorporates key components such as LangChain's ChatPrompt Template [7], HumanMessage, PromptTemplate, ConversationBuffer Memory, and LLMChain [8], creating an advanced solution for early detection and comprehensive support within the field of mental health. Additionally, the paper discusses the implementation of Streamlit to enhance the user experience and interaction with the chatbot.

The remainder of the paper is arranged accordingly. In Section II, LangChain and its important components are introduced. The proposed methodology for developing the whole architecture is described in Section III. Section IV provides an overview of Streamlit. Section V provides an illustration of sequential interaction of MindGuide chatbot and human. The conclusion is drawn in Section V.

## II. LANGCHAIN

LangChain, with its open-source essence, emerges as a promising solution, aiming to simplify the complex process of developing applications powered by large language models (LLMs). This framework though the rapid delivery of building blocks and pre-built chains for building large language model applications shows the easy way developers can do it.

* corresponding author - jkim72@kent.edu

LangChain helps us to unlock the ability to harness the LLM's immense potential in tasks such as document analysis, chatbot development, code analysis, and countless other applications. Whether your desire is to unlock deeper natural language understanding, enhance data, or circumvent language barriers through translation, LangChain is ready to provide the tools and programming support you need to do without it that it is not only difficult but also fresh for you. Its core functionalities encompass:

1. Context-Aware Capabilities: LangChain facilitates the development of applications that are inherently context-aware. This means that these applications can connect to a language model and draw from various sources of context, such as prompt instructions, a few-shot examples, or existing content, to ground their responses effectively.

2. Reasoning Abilities: LangChain equips applications with the capacity to reason effectively. By relying on a language model, these applications can make informed decisions about how to respond based on the provided context and determine the appropriate actions to take.

LangChain offers several key value propositions:

Modular Components: It provides abstractions that simplify working with language models, along with a comprehensive collection of implementations for each abstraction. These components are designed to be modular and user-friendly, making them useful whether you are utilizing the entire LangChain framework or not.

Off-the-Shelf Chains: LangChain offers pre-configured chains, which are structured assemblies of components tailored to accomplish specific high-level tasks. These pre-defined chains streamline the initial setup process and serve as an ideal starting point for your projects. The MindGuide Bot uses below components from LangChain.

### A. ChatModel

Within LangChain, a ChatModel is a specific kind of language model crafted to manage conversational interactions. Unlike traditional language models that take one string as input and generate a single string as output, ChatModels operate with a list of messages as input, generating a message as output.

Each message in the list has two parts: the content and the role. The content is the actual text or substance of the message, while the role denotes the role or source of the message (such as "User," "Assistant," "System," etc.).

This approach with ChatModels opens the door to more dynamic and interactive conversations with the language model. It empowers the creation of chatbot applications, customer support systems, or any other application involving multi-turn conversations. We utilized the ChatOpenAI ChatModel to create MindGuide chatbots specifically designed to function as mental health therapists. In our interaction with OpenAI, we opted for an OpenAI API key to engage with the ChatGpt3 turbo model and utilized a temperature value of 0.5. The steps to create an OpenAI API key are outlined [9].

### B. Message

In the context of LangChain, messages [10] refer to a list of messages that are used as input when interacting with a ChatModel. Each message in the list represents a specific turn or exchange in a conversation. Each message in the messages list typically consists of two components:

- content: This represents the actual text or content of the message. It can be a user query, a system instruction, or any other relevant information.
- role: This represents the role or source of the message. It defines who is speaking or generating the message. Common roles include "User", "Assistant", "System", or any other custom role you define.

The chat model interface is based around messages rather than raw text. The types of messages supported in LangChain are SystemMessage, HumanMessage, and AIMessage. SystemMessage is the ChatMessage coming from the system in its LangChain template as illustrated in Figure 1. Human Message is a ChatMessage coming from a human/user. AIMessage is a ChatMessage coming from an AI/assistant as illustrated in Figure 2.

> *You are a compassionate and experienced mental health therapist with a proven track record of helping patients overcome anxiety and other mental health challenges. Your primary objective is to support the patient in addressing their concerns and guiding them towards positive change. In this interactive therapy session, you will engage with the patient by asking open-ended questions, actively listening to their responses, and providing empathetic feedback. Your approach is collaborative, and you strive to create a safe and non-judgmental space for the patient to share their thoughts and feelings.*
>
> *As the patient shares their struggles, you will provide insightful guidance and evidence-based strategies tailored to their unique needs. You may also offer practical exercises or resources to help them manage their symptoms and improve their mental wellbeing. When necessary, you will gently redirect the conversation back to the patient's primary concerns related to anxiety, mental health, or family issues. This ensures that each session is productive and focused on addressing the most pressing issues. Throughout the session, you remain mindful of the patient's emotional state and adjust your approach accordingly.*
>
> *You recognize that everyone's journey is different, and that progress can be incremental.*
>
> *By building trust and fostering a strong therapeutic relationship, you empower the patient to take ownership of their growth and development. At the end of the session, you will summarize key points from your discussion, highlighting the patient's strengths and areas for improvement. Together, you will set achievable goals for future sessions, reinforcing a sense of hope and motivation. Your ultimate goal is to equip the patient with the tools and skills needed to navigate life's challenges with confidence and resilience.*

Figure 1. A System Message illustration

> *Welcome! to your therapy session. I'm here to listen, support, and guide you through any mental health challenges or concerns you may have. Please feel free to share what's on your mind, and we'll work together to address your needs. Remember, this is a safe and confidential space for you to express yourself. Let's begin when you're ready.*

Figure 2. An AIMessage illustration

*C. Prompt Template*

Prompt templates [10] allow you to structure input for LLMs. They provide a convenient way to format user inputs and provide instructions to generate responses. Prompt templates help ensure that the LLM understands the desired context and produces relevant outputs.

The prompt template classes in LangChain are built to make constructing prompts with dynamic inputs easier. Of these classes, the simplest is the PromptTemplate.

*D. Chain*

Chains [11] in LangChain refer to the combination of multiple components to achieve specific tasks. They provide a structured and modular approach to building language model applications. By combining different components, you can create chains that address various use cases and requirements. Here are some advantages of using chains:

- Modularity: Chains allow you to break down complex tasks into smaller, manageable components. Each component can be developed and tested independently, making it easier to maintain and update the application.
- Simplification: By combining components into a chain, you can simplify the overall implementation of your application. Chains abstract away the complexity of working with individual components, providing a higher-level interface for developers.
- Debugging: When an issue arises in your application, chains can help pinpoint the problematic component. By isolating the chain and testing each component individually, you can identify and troubleshoot any errors or unexpected behavior.
- Maintenance: Chains make it easier to update or replace specific components without affecting the entire application. If a new version of a component becomes available or if you want to switch to a differ.

To build a chain, you simply combine the desired components in the order they should be executed. Each component in the chain takes the output of the previous component as input, allowing for a seamless flow of data and interaction with the language model.

*E. Memory*

The ability to remember prior exchanges conversation is referred to as memory [12]. LangChain includes several programs for increasing system memory. These utilities can be used independently or as a part of a chain. We call this ability to store information about past interactions "memory". LangChain provides a lot of utilities for adding memory to a system. These utilities can be used by themselves or incorporated seamlessly into a chain.

A memory system must support two fundamental actions: reading and writing. Remember that each chain has some fundamental execution mechanism that requires specific inputs. Some of these inputs are provided directly by the user, while others may be retrieved from memory. In a single run, a chain will interact with its memory system twice.

1. A chain will READ from its memory system and augment the user inputs AFTER receiving the initial user inputs but BEFORE performing the core logic.
2. After running the basic logic but before providing the solution, a chain will WRITE the current run's inputs and outputs to memory so that they may be referred to in subsequent runs.

Any memory system's two primary design decisions are:

1. How state is stored ?
   Storing: List of chat messages: A history of all chat exchanges is behind each memory. Even if not all of these are immediately used, they must be preserved in some manner. A series of integrations for storing these conversation messages, ranging from in-memory lists to persistent databases, is a significant component of the LangChain memory module.
2. How state is queried ?
   Querying: Data structures and algorithms on top of chat messages: Keeping track of chat messages is a simple task. What is less obvious are the data structures and algorithms built on top of chat conversations to provide the most usable view of those chats.

A simple memory system may only return the most recent messages on each iteration. A slightly more complicated memory system may return a brief summary of the last K messages. A more complex system might extract entities from stored messages and only return information about entities that have been referenced in the current run. There are numerous sorts of memories. Each has its own set of parameters and return types and is helpful in a variety of situations.

*Memory Types:*

- *ConversationBufferMemory* allows for saving messages and then extracts the messages in a variable.
- *ConversationBufferWindowMemory* keeps a list of the interactions of the conversation over time. It only uses the last K interactions. This can be useful for keeping a sliding window of the most recent interactions, so the buffer does not get too large.

The MindGuide chatbot uses conversation buffer memory. This memory allows for storing messages and then extracts the messages in a variable.

III. ARCHITETURE

In crafting the architecture of the MindGuide app, each step is meticulously designed to create a seamless and effective user experience for those seeking mental health support. The user interface, built on Streamlit, sets the tone with a friendly and safe welcome. Users can jump in by typing their mental health questions, kicking off a series of interactions with the LangChain framework. This is where the magic happens – LangChain acts as the brain behind the chatbot, working through various components like chat message templates and a memory concept to create a personalized and responsive support system. Each step is broken down.

*Step 1.* User Interface: Developed using the Streamlit framework, the user interface welcomes users with a message explaining the role of the chatbot in providing mental health support. It assures users of a safe and confidential space to express their concerns.

*Step 2.* User Input - Prompt: Users can input mental health-related questions or seek advice by typing their queries into the input box integrated into the Streamlit interface.

*Step 3.* Data Transfer to LangChain: Implement the functionality that sends the user's input (question) as a chat prompt template to the LangChain framework. This input serves as the "human message prompt" template.

*Step 4.* LangChain Framework: In this phase, the LangChain framework serves as the backbone of the chatbot, where all the foundational components and building blocks are meticulously orchestrated. Here's a deeper dive into the critical elements of LangChain Processing:

- ChatMessage and Prompt Templates: Within LangChain, the chatbot's core communication infrastructure is established by creating ChatMessage and prompt templates for optimal chatbot engagement.
- LLMChain and LLM Model Interaction: To facilitate interactions with the large language model (LLM), a specialized component called LLMChain is constructed. The LLMChain acts as a conduit for managing the flow of conversation between the chatbot and the LLM model, in this case, GPT-4.
- The LLMChain handles both the user's queries and the chatbot's responses, allowing for a dynamic and coherent conversation flow.
- Chatmodel Class of LangChain: The LangChain framework leverages the Chatmodel class, a critical component for interfacing with the OpenAI model (GPT-4) for making requests to the language model and processing its responses, ensuring seamless communication between the chatbot and the AI model.
- Memory Concept: To enhance the chatbot's conversational capabilities and provide context-aware responses, LangChain incorporates a memory concept that allows the chatbot to retain and access information from past interactions within a session. The memory function enhances conversations by retaining user queries, preferences, and contextual details, thereby contributing to a more effective and personalized interaction. This way, it tailors responses based on the user's history throughout the session.

*Step 5.* Utilize the user's question as input to construct a chain of prompts that the large language model (in this case, GPT-4) will process.

*Step 6.* Model Response: Dispatch the constructed input chain to the GPT-4 model for natural language understanding and generation. The GPT-4 model generates a response based on the input and context.

*Step 7.* Response to Streamlit: Receive the response generated by the GPT-4 model and transmit it back to the Streamlit framework for display to the user.

*Step 8.* User Response Delivery: Present the model-generated response to the user, thereby delivering the mental health advice or information they sought.

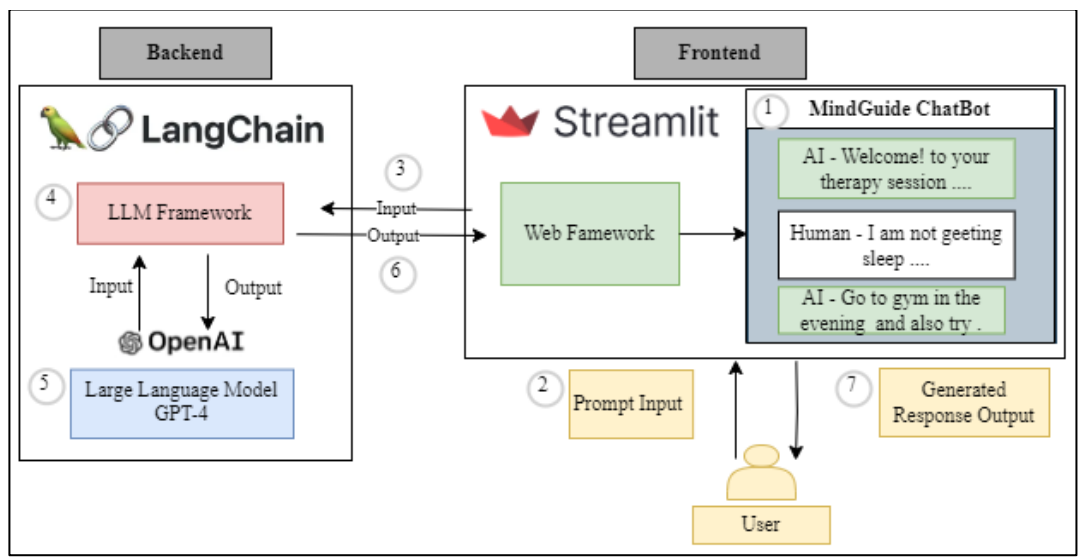

Figure 3. MindGuide Chatbot Architecture

## IV. STREAMLIT

Streamlit [13] is a faster way to build and share data apps. Streamlit turns data scripts into shareable web apps in minutes. Streamlit is an open-source Python library that simplifies the process of designing and sharing visually appealing web applications, particularly well-suited for applications involving machine learning and data science. Leveraging Streamlit's Python-based development approach, you can harness the power of Python to build a responsive and dynamic web application. This is advantageous for developers familiar with Python, as it allows for quick and efficient development.

## V. MINDGUIDE CHATBOT INTERACTION

The MindGuide Bot interaction is illustrated in Fig. 4, depicting the following key elements:

- Welcome screen interface with AI message and the initial human interaction with MindGuide Chatbot (Fig. 4a).
- MindGuide Chatbot's AI response to the human message, followed by the human's mental health question (Fig. 4b).
- MindGuide Chatbot's AI response to the subsequent human message, followed by another mental health question from the human (Fig. 4c).
- MindGuide Chatbot's AI response after analyzing the latest human message (Fig. 4d).

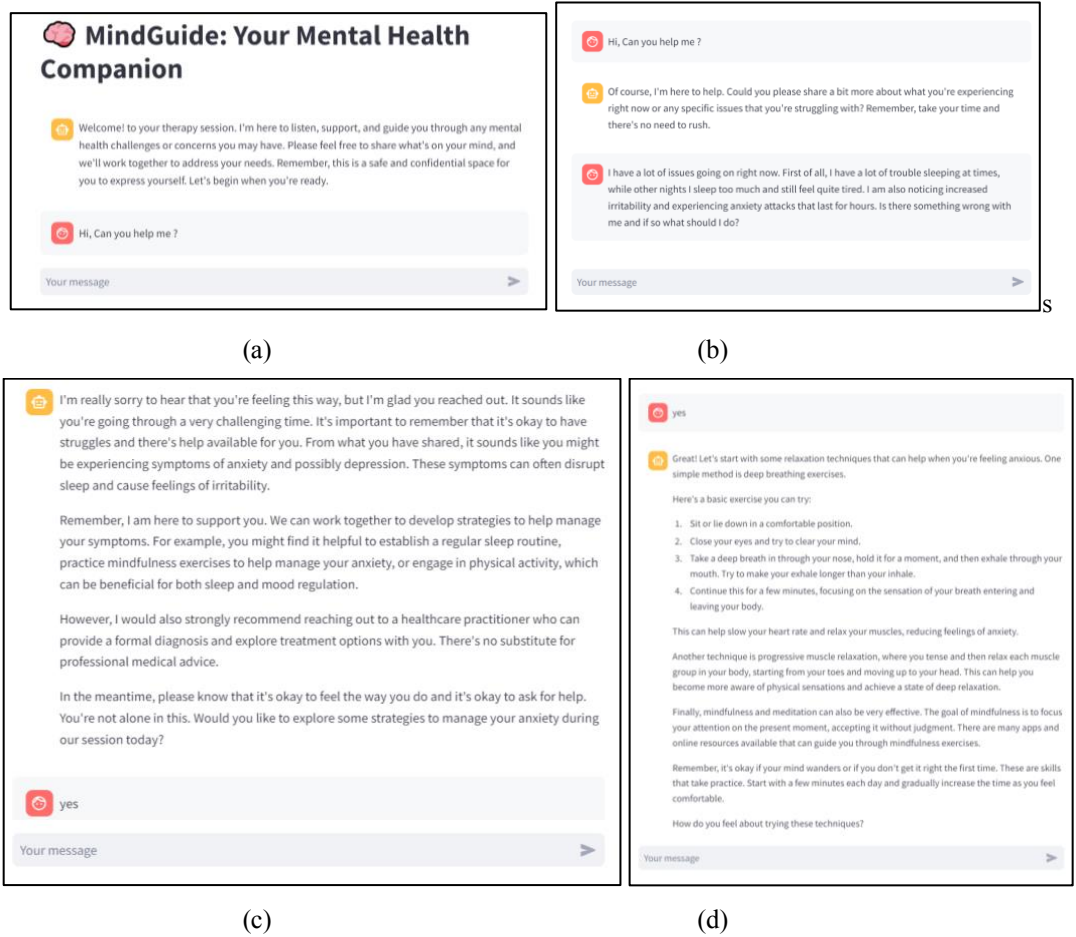

Figure 4. Sequential Interaction with MindGuide Chatbot - (a) Welcome screen and initial AI message, (b) AI response to the first human message and mental health question, (c) Subsequent AI response and continued interaction with another human mental health question, (d) AI response after analyzing the latest human message.

## VI. CONCLUSION

This paper employs the OpenAI chat model GPT-4 with a temperature setting of 0.5 to serve as an initial therapist, providing support for patients dealing with mental health issues such as depression and anxiety. MindGuide relies on the ChatOpenAI model from LangChain as its foundation, incorporating innovative features like ChatPrompt Template, Human Message Prompt Template, Conversation Buffer Memory, and LLMChain to proactively identify issues and deliver comprehensive assistance. In the next phase, we plan to enhance this chatbot further by implementing Retrieval-Augmented Generation (RAG) and incorporating embedding vectors for frequently asked questions related to mental health.